\newcommand{\vecri}{\mathrm{\mathbf{r_i}}}
\newcommand{\vecrj}{\mathrm{\mathbf{r_j}}}
\newcommand{\gw}{g_{\omega}}
\newcommand{\alpi}{a_i}

\newcommand{\Einci}{E^{\mathrm{inc}}(x_i)}
\newcommand{\Einc}{E^{\mathrm{inc}}}

\newcommand{\myham}{\mathcal{M}}
\newcommand{\offdiag}{D}
\newcommand{\xij}{k|i-j|}
\newcommand{\pdf}{\mathcal{P}}

\newcommand{\be}{\begin{equation}}
\newcommand{\en}{\end{equation}}

\newcommand{\re}{\mathrm{Re}\,}
\newcommand{\im}{\mathrm{Im}\,}
\newcommand{\fa}{f(\alpha)}
\newcommand{\fpa}{f'(\alpha)}
\newcommand{\tauq}{\tau(q)}
\newcommand{\taupq}{\tau'(q)}
\newcommand{\ket}[1]{\left| #1 \right>} 
\newcommand{\braket}[2]
{\left< #1 \vphantom{#2} \right| \left. #2 \vphantom{#1}\right>} 

\documentclass[aps,prb,preprint]{revtex4}
\usepackage{amsmath,array,graphicx}
\begin{document}

\title{Critical scaling of polarization waves on a heterogeneous chain of resonators}
\author{Sanli Faez}
\email{faez@amolf.nl}
\affiliation {FOM Institute for Atomic and Molecular Physics AMOLF, Science Park 104,
1098 XG Amsterdam, The Netherlands}
\author{Ad Lagendijk}
\affiliation {FOM Institute for Atomic and Molecular Physics AMOLF, Science Park 104,
1098 XG Amsterdam, The Netherlands}
\author{Alexander Ossipov}
\affiliation{School of Mathematical Sciences, University of Nottingham,
Nottingham NG72RD, United Kingdom}
\begin{abstract}
The intensity distribution of electromagnetic polar waves in a chain of
near-resonant weakly-coupled scatterers is investigated theoretically
and supported by a numerical analysis.
Critical scaling behavior is discovered for part
of the eigenvalue spectrum due to the disorder-induced Anderson
transition. This localization transition (in a formally
one-dimensional system) is attributed
to the long-range dipole-dipole interaction, which decays
inverse linearly with distance for polarization perpendicular to the chain.
For polarization parallel to the chain, with
inverse squared long range coupling,
all eigenmodes are shown to be localized. A
comparison with the results for Hermitian power-law
banded random matrices and other intermediate models is presented.
This comparison reveals the significance of non-Hermiticity
of the model and the periodic modulation of the coupling.
\end{abstract}



\maketitle
\section{Introduction}

Collective excitations of nanoparticle composites have shown
promising applications for
sensing, nonlinear spectroscopy, and photonic circuits. Among
these applications, transport of electromagnetic signal along an assembly of
metallic nanoparticles has been the subject of intensive
research in recent years. It has shown promising
applications in integrated photonics\cite{alu_theory_2006}
and sensing\cite{polman_2008}. By the nature of
their fabrication, disorder is inevitable in these artificial structures and
therefore must be considered accordingly.

In this article, we make a connection between the photonic transport in these novel
physical structures and the Anderson localization transition: a
fundamental phenomenon which emerges in various fields such as
condensed matter physics~\cite{evers_anderson_2008}, cold gases in optical
lattices~\cite{ingusciu_phystoday} and
classical waves in random media~\cite{lagendijk_phystoday}. We argue how the polar
excitations in a chain of resonators can show \emph{critical} scaling
behavior. This criticality plays a major role in understanding the
underlying phase transition phenomena.

Anderson localization in electronic systems is extensively
studied in the form of the Anderson tight-binding Hamiltonian
with on-site disorder\cite{kramer_finite_2010}. For this model all states are
exponentially-localized in one and two dimensions. In three dimensional space
there exists a metal-insulator transition, as suggested by
the single parameter scaling theory. At the
Anderson transition, wavefunctions show critical statistical behavior and their
spatial structure are
multifractal~\cite{evers_anderson_2008,wegner_inverse_1980,aoki_critical_1983}.

Levitov~\cite{levitov_absence_1989,levitov_delocalization_1990}
has studied the effect of long-range
interaction, in the form of $D_{ij}\propto|\vecri-\vecrj|^{-\mu}$,
on the the localization of vibrational excitations of a disordered lattice.
He showed that for $\mu > d$, where $d$ is the dimensionality of space,
all states remain localized. For $\mu < d$ all states
escape localization due to a diverging number of
resonances between spectrally apart energy levels.
For $\mu = d$, delocalization is weak and
states are critical.

Transition from localized to extended eigenstates has been
shown by Mirlin et.
al.~\cite{mirlin_transitionlocalized_1996}
in the ensemble of power-law random banded matrices (PLRBM). In these
Hermitian random matrices the off-diagonal elements decay as
$\langle |H_{ij}|^2 \rangle = (b/|i-j|)^{2\mu}$ for $|i-j|>b$,
where $b$ is called the bandwidth. Similar to Levitov's model all
eigenvectors are localized for $\mu > 1$, the tight-binding limit, and
are extended (metallic) for $\mu < 1$, as for conventional Wigner-Dyson random
matrices. At $\mu = 1$, eigenstates show critical statistics for
any width of the band.

A localization transition is also shown for a non-Hermitian
disordered tight-binding model by Hatano and
Nelson~\cite{hatano_localization_1996}. Their model was motivated by
its application to a special mapping of flux lines in
certain superconductors to a bosonic system with a random potential.

The physical system considered in this report,
is fully described by a class of complex-symmetric coupling matrices.
The eigenvectors of these matrices describe the excitation ``modes'' of a
linear chain of point-like scatterers. These scatterers
are driven close to resonance and are coupled to each other
through long-range
dipole-dipole interaction.

We have studied this system in two cases of weak and strong coupling.
By studying the scaling behavior of eigenstates in the weak coupling regime,
we show that
for transverse magnetic (TM) polarization parallel to the chain direction, all the
states are localized. For
the transverse electromagnetic (TEM) polarization in this regime,
some of the states are critically
extended and their scaling is described by a multifractal spectrum.
We analytically derive a perturbation expression for this multifractal spectrum in the
limit of weak coupling corresponding to very strong
disorder. In the strong coupling regime, we show some
numerical evidence that the intensity distribution follows a
mixed phase of localized and extended statistics.

We have extensively compared the scaling behavior of this physical
system with several hypothetic Hermitian
and non-Hermitian matrix ensembles. This comparison proves the strong influence of the
phase periodicity in the coupling terms. For example, the
Levitov matrices with $\mu=1$ are no longer critical, but localizing, if
the off-diagonal terms are non-random and have a periodic phase relation.
On the other hand, the observed critical behavior of TEM
polar eigenmodes disappear if the interaction phase factor
is randomized.

Our findings provide a clear and universal framework for excitation properties of
an important building block in modern photonics. On a broader perspective our model has
significant resemblance with  the other important classes
of Hamiltonians, which are used for describing several transport phenomena in
mesoscopic systems. Since our model has an exact
correspondence to a
real physical system, it will pave the way
for experimental investigation of several theoretical
findings, which up to now were bound to the limitations of numerical simulation.

\section{The model}
For describing the chain of resonators, we use the dipole
approximation for each of the scatterers and the full
dyadic on-shell Green function for their interaction.
This model is previously used for describing collective plasmon excitations of
metallic nanoparticles on a line or a plane for
periodic~\cite{weber_propagation_2004,koenderink_complex_2006},
aperiodic~\cite{forestiere_role_2009} and
disordered configurations~\cite{markel_anderson_2006}.
In particular Markel and Sarychev
have reported signatures of localization in a
chain of point-like
scatterers~\cite{markel_propagation_2007}.

The presence of an Anderson transition, its critical behavior, and the detailed
statistics of localized or delocalized modes in such a
system has not yet been studied. In the following we will argue
and show analytically that a disorder mediated delocalization
transition can happen for polarization perpendicular to the chain
direction (TEM modes), while for polarization parallel to
the chain direction (TM modes), all eigenstates are localized in a long enough chain.
We will present our results using a well-established
statistical framework of probability density function (PDF)
of eigenmode intensities and the scaling of generalized
inverse participation ratios (GIPR).

\subsection{Dipole chain model}\label{sec:dipolechainmodel}
We consider a linear array of $L$ equally spaced polarizable
isotropic particles with an interparticle distance of $s$. The
size of the particles are considered small enough,
relative to both $s$ and the excitation wavelength $\lambda\equiv 2\pi c/\omega$, for
a point-dipole approximation to be valid. With these
considerations, TEM and TM modes are decoupled from each other.
For the stationary response, oscillating with constant frequency $\omega$, the dipole
moments of particles $p_i\equiv \mathbf{\hat{u}}.\mathrm{\mathbf{p}}_i$
projected on each mode, are the solutions to the
following homogeneous set of linear equations
\begin{equation}\label{eq:coupleddipoles}
p_i(x_i) = \alpi(\omega) \left[\Einci+\sum_{j\neq
i}\gw\left(|x_i-x_j|\right)p_j(x_j)\right],
\end{equation}
where $\alpi$ is the polarizability of the $i$th particle and
$e^{\imath\omega t}\Einci$ is the
incident electric field at its position projected on the specific Cartesian
coordinate of the mode, $\mathbf{\hat{u}}$. The free
space Green function $\gw$ should be replaced by the proper expressions for
TEM($\perp$) and TM($\parallel$) modes, which are given by
\begin{eqnarray}\label{eq:freegreenfunction}
\gw^{\perp}(x)&=&\frac{1}{4 \pi \epsilon}\left(\frac{\omega^2}{c^2 x}
+\frac{\imath\omega}{c x^2}-\frac{1}{x^3}\right)
e^{\imath\omega x/c},\\
\gw^{\parallel}(x)&=&\frac{-1}{2 \pi \epsilon}\left(
\frac{\imath\omega}{c x^2}-\frac{1}{x^3}\right)
e^{\imath\omega x/c},
\end{eqnarray}

Equation~(\ref{eq:coupleddipoles}) can be represented in its
matrix form $\myham \ket{p}= \ket{\Einc}$ where
\begin{equation}\label{eq:defineh}
\myham_{ij}=\delta_{ij}\alpi^{-1}+(1-\delta_{ij})\gw(|x_i-x_j|).
\end{equation}
The explicit frequency dependence of
$\alpi$ is dropped, since we consider only monochromatic
excitations in this article. The matrix $\myham$ is a complex and
symmetric matrix, the inverse of which gives the polarization response
of the system to an arbitrary excitation; $\ket{p}= \myham^{-1}
\ket{\Einc}$. In fact $\myham^{-1}$ is the so-called
$t$-matrix~\cite{de_vries_point_1998}
of the chain specified on the lattice points.
Since $\myham$ is non-Hermitian, its eigenvalues are complex.
However,
the eigenvectors form a complete (bi-orthogonal) basis, unless the matrix
is defective. Defective matrices form a subset of measure zero
for a randomly generated ensemble. No defective matrix has
been encountered in this research.
The orthogonality condition is set by the quasi-scalar product of
each two eigenvectors:
\begin{equation}\label{eq:orthogonality}
\braket{\bar{\psi}_m}{\psi_n}\equiv\sum_i{{\psi}_m(x_i){\psi}_n(x_i)}=0,
\end{equation}
where $\ket{\psi_n}$ is a right eigenvector of $\myham$;
$\myham\ket{\psi_n}=\varepsilon_n\ket{\psi_n}$.
The eigenvectors are normalized
to unity: $\braket{{\psi}_n}{\psi_n}=1$.
The quasi-scalar product of an eigenvector with
itself is a non-zero complex number for non-defective matrices.

Under the stated assumptions, the polarization response to
an incident field can be obtained from the decomposition
\begin{equation}\label{eq:polarizationdecomposition}
\ket{p}=\sum_n\frac{\ket{\psi_n}\braket{\bar{\psi}_n}{\Einc}}
{\varepsilon_n\braket{\bar{\psi}_n}{{\psi}_n}}.
\end{equation}
A null eigenvalue points to a collective resonance
of the system and the corresponding eigenvector is the
most bound (guided) mode with the highest polarizability.

\subsection{Resonant point scatterer}
A simple and yet general model for the dipolar polarizability of a
point scatterer that conserves energy~\cite{de_vries_point_1998} is given by
\begin{equation}\label{eq:polarisability}
\frac{1}{a}=\frac{1}{4 \pi \epsilon}\left(\frac{1}{a^\mathrm{D}}-
\frac{2\imath\omega^3}{3c^3}\right),
\end{equation}
where the last term in Eq.~(\ref{eq:polarisability})
is the first non-vanishing radiative correction that
fullfils the optical theorem.
The quasistatic polarizability $a^\mathrm{D}$
depends on the particle shape and
its material properties. For a Lorentzian resonance around
$\omega_R$
\begin{equation}\label{eq:drudepolarisability}
\frac{1}{a^\mathrm{D}}=\frac{A}{V}\left(1-\frac{\omega^2+\imath\gamma\omega}{\omega_R^2}\right),
\end{equation}
where $V$ is the volume of the scatterer, $\gamma$ is the Ohmic damping factor
and $A$ is a constant that depends only on the geometry of the scatterer.
For elastic scatterers
$a^\mathrm{D}$ is real-valued
and diverges on resonance.

\subsection{Dimensionless formulation}\label{sec:dimless}
To study the properties of the coupling
matrix~(\ref{eq:defineh}) both theoretically and numerically,
we rewrite it in terms of dimensionless quantities by dividing
all the length dimensions by the interparticle distance $s$ and
multiplying the unit of polarizability by $4 \pi \epsilon k^3$, where
$k=\omega/c$. For the cases
considered in this article, we also neglect the Ohmic damping of scatterers
and hence the imaginary part on the diagonal of the matrix is given by the radiative
damping term in Eq.~(\ref{eq:polarisability}).

Based on definition~(\ref{eq:defineh}) two distinct types of disorder can be
considered for the system under investigation. Pure off-diagonal
disorder is caused by the variation in the inter-particle
spacing considering identical scatterers.
The contrary case of diagonal disorder
applies when the particles are positioned periodically but
have inhomogeneous shapes or different resonance frequencies.
For the sake of brevity, we limit our discussion to the case of pure diagonal disorder.
All the techniques used in this article are also applicable in presence of
off-diagonal disorder.

In the units described before, the off-diagonal elements of $\myham$ are written as
\begin{eqnarray}
\offdiag^{\perp}_{i\neq j}&\equiv& \left(-\frac{1}{\xij}-\frac{\imath}{(\xij)^{2}}
+\frac{1}{(\xij)^{3}} \right) e^{\imath \xij},\label{eq:dimlesste}\\
\offdiag^{\parallel}_{i\neq j}&\equiv& 2\left(\frac{\imath}{ (\xij)^2} - \frac{1}{(\xij)^3}
\right) e^{\imath \xij}\label{eq:dimlesstm},
\end{eqnarray}
for TEM and TM excitations, respectively.

Since the Ohmic damping is absent and the lowest order radiation damping is
independent of the particle geometry, the diagonal elements
are inhomogeneous only in their real parts.
We choose the real part from the
the set of random numbers $U(-W/2,W/2)$ which has a box probability distribution
around zero with a width $W$. The imaginary part of the diagonal elements
is constant in these units and equals $-2\imath/3$. Considering the linear
dependence of the inverse of
polarizability~(\ref{eq:drudepolarisability}) on the particle
volume and detuning from resonance frequency, realizing a
uniform distribution is practical.

\subsection{Hypothetic models}\label{sec:models}
The results of the perturbation approximation disagrees with
some of the trends observed in our numerical results for weakly coupled systems. To
shed light on the origin of these observations, we have
performed similar statistical analysis on extra
hypothetical models. In these four models, step by step, we transform
our model for TEM excitation to
an ensemble of orthogonal random matrices, for which
extensive results have been reported in the literature
(see Ref.~[\onlinecite{evers_anderson_2008}] for a recent review). In
all these ensembles the diagonal elements are real random
numbers selected from the set $U(-W/2,W/2)$. The distinction is
in the off-diagonal elements which are defined as follows:
\begin{enumerate}
\item[H0,] The matrices in this model are orthogonal and
they are the closest to the frequently used PLRBM ensemble.
The offdiagonal elements are random real numbers given by
\begin{eqnarray}\label{eq:modelh0}
\offdiag^{\mathrm{H0}}_{i\neq j}&\equiv&\frac{h_{ij}}{\xij},
\end{eqnarray}
where $h_{ij}$ is a randomly chosen from $U(-1,1)$; i.e. uniformly distributed in [-1,1].
\item[H1,] These matrices are the Hermitian counterpart of the
TEM coupling matrix with a randomized phase factor for each element:
\begin{eqnarray}\label{eq:modelh1}
\offdiag^{\mathrm{H1}}_{i <
j}&\equiv&\offdiag^{\perp}_{ij}e^{\imath\phi_{ij}},\\ \nonumber
\offdiag^{\mathrm{H1}}_{i > j}&\equiv&\bar{\offdiag}^{\perp}_{ij} e^{-\imath\phi_{ji}},
\end{eqnarray}
where $\phi_{i<j}$ is a random number from $U(-\pi,\pi)$.
\item[C1,] This ensemble of complex-symmetric matrices
resemble the TEM model with a randomized coupling phase.
\begin{eqnarray}\label{eq:modelc1}
\offdiag^{\mathrm{C1}}_{ij}&\equiv&\offdiag^{\perp}_{ij}e^{\imath\phi_{ij}},
\end{eqnarray}
where $\phi_{ij}\equiv\phi_{ji}$ are random numbers from $U(-\pi,\pi)$.
\item[H2,] This model is based on the Hermitian form of TEM
interaction and the phase factor is kept periodically varying.
\begin{eqnarray}\label{eq:modelh2}
\offdiag^{\mathrm{H2}}_{i < j}&\equiv&\offdiag^{\perp}_{ij},\\ \nonumber
\offdiag^{\mathrm{H2}}_{i > j}&\equiv&\bar{\offdiag}^{\perp}_{ij}.
\end{eqnarray}
\end{enumerate}

\section{Analytic results}
Decomposition~(\ref{eq:polarizationdecomposition}) relates
the overall statistical behavior of the system to the
properties of the eigenmodes and their corresponding
eigenvalues. The dipole chain is an open system and
the excitations are subject to radiation losses, which lead to
the exponential decay of a mode. Therefor it is not possible
to distinguish between disorder and loss origins of localization
only based on the spatial extent of a mode. For these types
of systems, statistical analysis has shown to be the only
unambiguous method of studying Anderson transition.
Therefor we study the scaling behavior. For this analysis,
based on the eigenvectors in the position basis,
two important indicators are considered:
1-the probability distribution function (PDF) of the wavefunction intensities and
2-the generalized inverse participation ratios (GIPR).

The PDF is more easily accessible in experiments~\cite{krachmalnicoff_fluctuations_2010}.
For numerical analysis, it has
proven to be an accurate tool for measuring the scaling
exponent in a finite size
system~\cite{rodriguez_multifractal_2009} and extracting
the critical exponent from finite size scaling
analysis~\cite{rodriguez_critical_2010}.
With the parametrization $\pdf(\tilde{\alpha};W,L,b)$ the PDF is
sufficient for characterizing an Anderson transition. Here,
$\tilde{\alpha} \equiv \ln I_B / \ln (b/L)$
with $I_B \equiv \sum_{i=1}^{b}\left|{\psi}_n(x_i)\right|^2$
the integrated intensity over any box selection of length
$b$. The effective disorder strength is parameterized by $W$, but the
exact definition depends on the model.
Criticality of eigenfunctions demands the scale invariance
of the PDF. It means that the functional form of $\pdf$
does not change with system size for a fixed
$b/L$. Away from the
transition point, the maximum of the PDF, $\tilde{\alpha}_m$, exhibits
finite size scaling behavior~\cite{rodriguez_critical_2010}.
This maximum shifts to higher(lower) values at
the localized(extended) side of the transition.

Another widely used set of quantities for evaluating the
scaling exponents is the set of GIPR, which are proportional to the moments of the PDF.
For each wavefunction GIPR are defined as
\begin{equation}\label{eq:gipr}
P_{q}(\{{\psi}_n\}) \equiv \sum^L_{i=1}\left|{\psi}_n(x_i)\right|^{2q}.
\end{equation}
At criticality, the ensemble averaged GIPR, $\langle P_q\rangle$, scales
anomalously with the length $L$ as
\begin{equation}\label{eq:dq}
\langle P_q\rangle\sim L^{-d_q(q-1)},
\end{equation}
where $d_q$ is called the anomalous dimension. For
multifractal wavefunctions, which are characteristic of
Anderson transitions
$d_q$ is a continuous function of
$q$. From the definition, $P_1=1$ and $P_0= L$. In practice,
the GIPR can also be evaluated by
box-scaling for a single system size, given a
large enough sample~\cite{vasquez_multifractal_2008}.

\subsection{Perturbation results for the weak coupling
regime}\label{sec:perturbation}

In the regime of weak coupling $Wk\gg1$ the off-diagonal
matrix elements of the Hamiltonian are small compared to the diagonal ones.
Therefore the moments of the eigenfunctions can be computed perturbatively
using the method of the virial expansion
\cite{levitov_delocalization_1990, mirlin_evers_2000, yevtushenko_virial_2003, kravtsov_2010}.
To this end we generalize the route suggested in Ref.~[\onlinecite{fyodorov_2009}]
to the case of the non-Hermitian random matrices.

By using this perturbation analysis, we find that TEM eigenfunctions
scale critically with the
length of the system. The criticality is set by the
inverse linear interaction term
in Eq.~(\ref{eq:dimlesste}) which dominates at large distances. In the weak-coupling regime,
the set of multifractal exponents can be explicitly calculated. The result is different
from the universal one found for all critical models with Hermitian random matrices \cite{fyodorov_2009}
and is given by
\begin{eqnarray}\label{eq:dq}
d_q&=&\frac{2 c_0(q)}{W k (q-1)},\quad q>\frac{1}{2}.
\end{eqnarray}

The detailed derivation of this result as well as an explicit expression for $c_0(q)$ are
presented in Appendix~\ref{app:dq}. The corresponding result for the  orthogonal
 matrices reads
\begin{eqnarray}\label{eq:dqhermitian}
d_q&=&\frac{4 \sqrt{\pi}\Gamma(q-1/2)}{W k \Gamma(q)},\quad q>\frac{1}{2}.
\end{eqnarray}

If a similar analysis is performed on the TM eigenfunction,
the GIPR converge at large system sizes implying that
the eigenfunctions are localized. This is due to the $r^{-2}$
behavior of the coupling at large distances.

In the following section, an extensive comparison is made between
these analytical expressions and the numerical simulations.

\section{Numerical results}

By direct diagonalization of a
large ensemble of matrices, we have studied the PDF and GIPR scaling of the
eigenfunctions of matrices from all the models introduced in the previous
sections. Several values of disorder strength $W$
and carrier wavenumber $k$ are considered for matrices with
sizes from $L=2^7$ to $2^{12}$.
Each matrix is numerically diagonalized with MATLAB using the ZGGEV
algorithm. The number of analyzed eigenfunctions for each
set of parameters is around $10^4$.
Computation time for diagonalization of the largest matrix is
20 minutes on a PC.

\subsection{Spectrum of the homogeneous chain}
We start by analyzing the spectrum of the homogenous
chain on resonance (W=0) where all the diagonal element are given by
$\myham_{ii}=0-2\imath/3$. Typical spectra for TEM and TM excitations
are shown in Figures~\ref{fig:spectrate}(a) and \ref{fig:spectratm}(a) for $k=1$.

For $k<1.4$, the TEM eigenvalues are divided into almost-real and complex subsets.
The almost-real ($\im{\varepsilon} \ll \re{\varepsilon}$) subset corresponds
to subradiative (bound)
eigenstates. These eigenstates have a wavelength shorter than the
free space propagation\cite{weber_propagation_2004,koenderink_complex_2006}
and cannot couple to the outgoing
radiation, except at the two ends of the chain. The
eigenmodes corresponding to complex eigenvalues
($\im{\varepsilon} \sim \re{\varepsilon}$)
are superradiative.
For these modes a constructive interference in the far-field enhances the
scattering from each particle in comparison with the an isolated one.

From the form of
expansion~(\ref{eq:polarizationdecomposition}) it is clear
the eigenstates with (close to) zero eigenvalues will
dominate the response of the system to external excitation.
However, different regions in the spectrum can be experimentally probed by
two approaches: Firstly, by changing the lattice spacing, or secondly, by detuning from the
resonance frequency, which will add a
constant real number to the diagonal of the interaction
matrix~$\myham$. Close to the resonance this number is
linearly proportional to the frequency variation.
This shift results in driving a different
collective excitation, which has obtained the closest eigenvalue to
the origin of the complex plane.

\subsection{The effect of disorder}
As mentioned before, disorder is introduced to the system
by adding random numbers from the interval $[-W/2,W/2]$ to
the diagonal of $\myham$. With this setting, the parameter
space has two coordinates $W$ and $k$, and $g=(Wk)^{-1}$ is the coupling parameter.
The weak and strong coupling regimes correspond to $g\ll1$ and
$g\approx\mathcal{O}(1)$ respectively. For $k<0.5$ the short range behavior
is dominated by the quasi-static part of the interaction. The
eigenstates of the disordered chain are thus exponentially
decaying --similar to localized states.
Since we are mainly interested in the critical behavior of
eigenfunctions we focus on the region with $0.5<k<3$.

\subsubsection{Small disorder}
\begin{figure}[htb]
  \includegraphics[width=8cm]{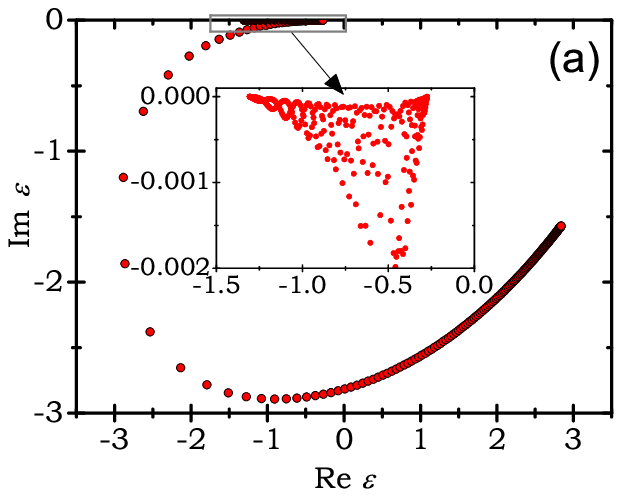}\\
    \includegraphics[width=8cm]{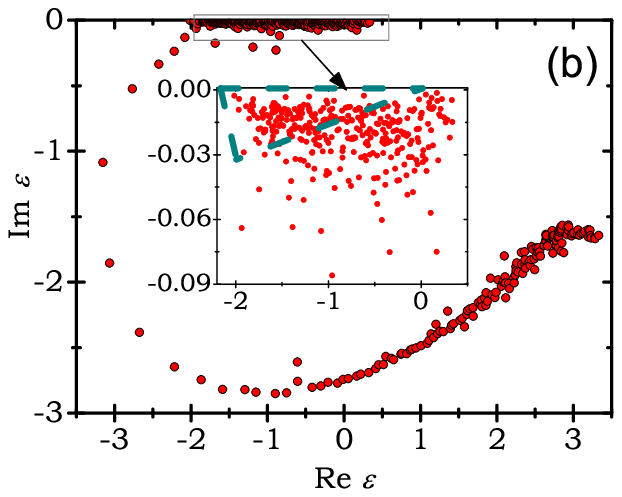}\\
  \includegraphics[width=8cm]{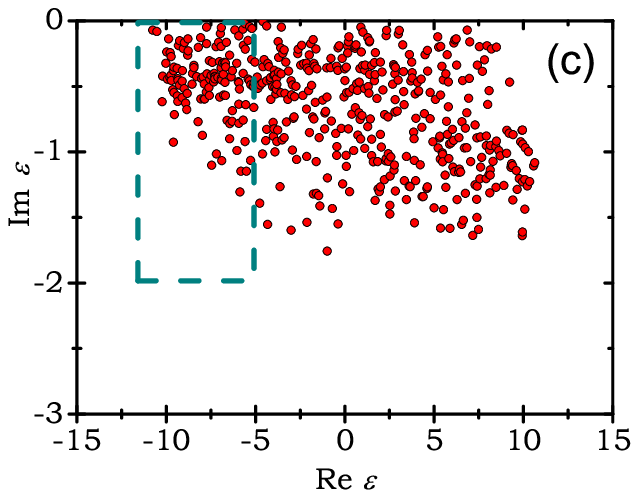}
  \caption{\label{fig:spectrate}Complex valued spectrum of the TEM interaction
  matrix~(\ref{eq:dimlesste}) with $k=1$ for (a) homogeneous ($W=0$) and
  disordered in regimes of (b) strong ($W=2$) and (c) weak ($W=20$)
  coupling. The dashed square in (c) shows the
  region where the corresponding TEM eigenmodes are
  scaling critically. The eigenmodes corresponding to the eigenvalues in
  the dashed regions are selected for further statistical analysis.  }
\end{figure}

\begin{figure}[htb]
  \includegraphics[width=8cm]{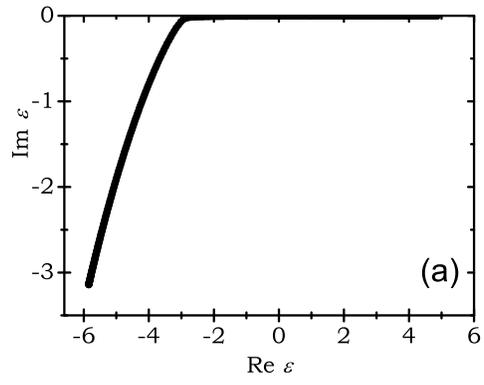}\\
  \includegraphics[width=8cm]{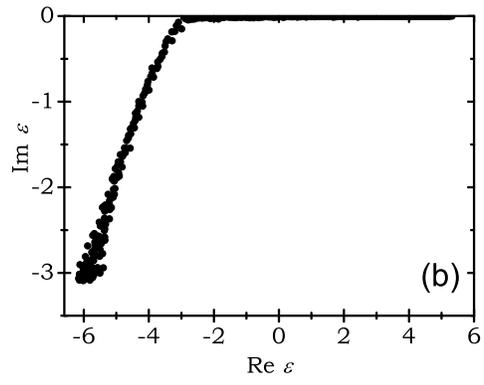}\\
  \includegraphics[width=8cm]{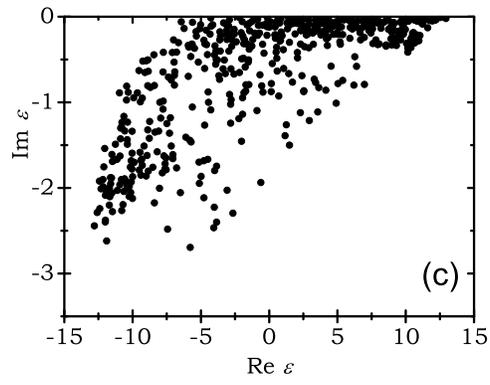}
  \caption{\label{fig:spectratm} Same as Fig.~\ref{fig:spectrate}
  for TM polarization.}
\end{figure}

In the intermediate and strong coupling regime, the
spectrum of the disordered matrix keeps the overall form of the
homogeneous case (where the real part of the diagonal is
zero), as can be seen in Figures~\ref{fig:spectrate}(b)
and~\ref{fig:spectratm}(b).

At the sub-radiative band-edge of the TEM spectrum, modes of different
nature mix due to disorder.
This region is magnified in the inset of Fig.~\ref{fig:spectrate}(b).
The eigenmodes corresponding to this region are of hybrid character.
They consist of
separate localization centers that are coupled via extended
tails of considerable weight.
The typical size of each localized section is longer than the
interparticle spacing. Similar modes have been observed in a quasi-static
investigation of two-dimensional planar composites\cite{stockman_localization_2001}. For one-dimensional
systems they are sometimes called necklace states in the
literature~\cite{pendry_quasi-extended_1987,bertolotti_optical_2005}.
Heuristically, this behavior can be attributed to the disorder induced mixing
of sub-radiative and super-radiative modes which have closeby eigenvalues
in the complex plain. Further evidence for this mixed behavior will be
later discussed based on the shape of PDF in section~\ref{sec:strongcoupling}.

For TM polarization all of the eigenstates become exponentially localized
with power-law decaying tails. The localization
length increases towards the band center. Therefore, in a chain with finite length, one
will see two crossovers in the first Brilluoin zone, from localized
to extended and back. However, the nature of localization seems to be different
at the two ends. The subradiative modes ($\im{\varepsilon} \ll \re{\varepsilon}$)
are localizaed due to interference effects
similar to the Anderson localization while the superradiative
modes ($\im{\varepsilon} \sim \re{\varepsilon}$) are
localized by radiation losses.
These two crossover regions
eventually approach each other and disappear as the amount of
disorder is increased, leading to a fully localized spectrum of eigenmodes.
The spectral behavior is more complicated for higher
wavenumbers with $k>\pi$ but a discussion on that is further than the scope
of this article.

\subsubsection{Large disorder}
\begin{figure}[htb]
  \includegraphics[width=8cm]{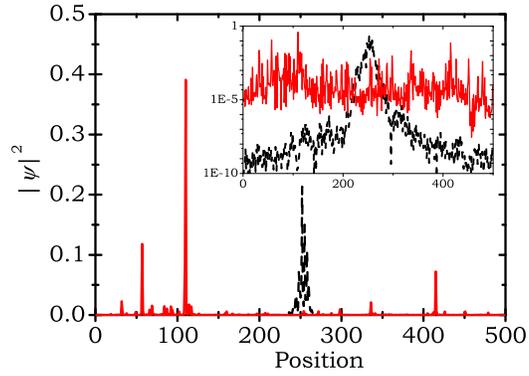}
  \caption{\label{fig:eigenmodes}Typical TEM critical
  (red solid line)
  and TM localized (black dashed line) eigenvectors of matrices defined in section~\ref{sec:dimless}
  with $k=1$ and $W=10$. The inset shows the same plot in logarithmic scale.}
\end{figure}

In the weak coupling regime, the matrix is almost diagonal
and thus the eigenvalues just follow the distribution of the
diagonal elements. Typical eigenstates are shown in Fig.~\ref{fig:eigenmodes}.
As will be shown later, for this regime, all the eigenstates for
TEM and TM are localized (since the coupling is weak)
except for a band (about 20\% width) of TEM eigenstates with the most negative
real part of their eigenvalues.
The states in this band show multifractal (critically extended)
behavior for any arbitrarily weak coupling. Existence of these states is
one of the major results of this investigation and their statistical
analysis is the main subject of interest in the rest of this report.

The multifractality of eigenfunction in the weak coupling regime
is inline with the prediction of the virial expansion result~(\ref{eq:dq}).
However our theory
cannot describe why only a part of the TEM eigenstates are critically extended
and the rest of them are localized, according to the numerical results.

%

\subsection{Scaling behavior of PDF}
The scaling of PDF is an effective tool for analyzing the
localized to extended transition in sample with finite
length~\cite{rodriguez_multifractal_2009,rodriguez_critical_2010,schubert_distribution_2010}.
We also use this statistical indicator to distinguish the regions
of critical scaling. Only those eigenmodes for which
their scaled
PDF for different system sizes overlap are critical. For the
wavefunctions that fulfill this criteria, the scaling of
GIPR is analyzed. This second analysis confirms the
presence of critical behavior by checking the the power-law
scaling behavior. The logarithmic slope gives the multifractal
dimensions.
We have preformed
extensive survey of the size-scaling behavior of PDF over the
$W,k$ space with $~10^4$ eigenfunctions for each configuration.

\begin{figure}[htb]
  \includegraphics[width=8cm]{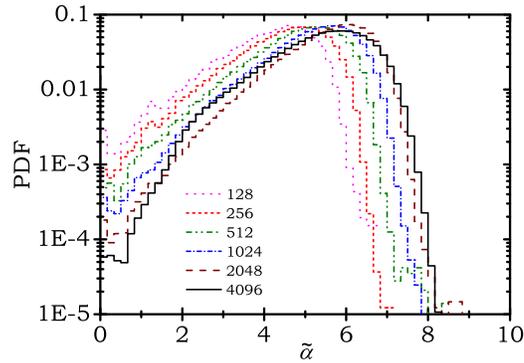}
  \caption{\label{fig:pdftm}Scaling of PDF for TM
  eigenmodes for different lengths of
  the chain and correspondingly scaled
  box sizes of $b=2^{-6}\times L$. Different line types (and
  colors) correspond to different system sizes as indicated
  by the legend. $W=30$ and $k=1$.
  The shift of the peak to the larger values of $\tilde{\alpha}$
  indicates that the eigenmodes are localized.}
\end{figure}

For each system size the scaled PDF is approximated by a
histogram $\mathcal{P}(\ln I_B/\ln (b/L))$ over the sampled eigenfunctions
These histograms are shown in
Figures~\ref{fig:pdftm} to \ref{fig:pdfhyp} for different models.
The shift of the peak of the distribution toward larger values
(higher density of darker points) by an increase in the system size
is a signature of eigenmode localization. A shift in the opposite
direction towards a Guassian distribution with a peak at
$\tilde{\alpha}_m=d=1$ is characteristic of the extended states. Overlap
of these histograms signifies the critical behavior of the
eigenmodes.

\subsubsection{TM and TEM modes in the weak coupling regime}
The typical scaling of PDF for TM modes is shown in
Fig.~\ref{fig:pdftm}. It clearly reveals the localized behavior of
these eigenfunctions. This is the generic behavior observed
for these modes at any point in the parameter space. This result is
in agreement with the Levitov's prediction, since the coupling is
decaying as $r^{-2}$.
Localization in disordered one dimensional systems has already been studied
extensively and we do not discuss it
further in this article.

\begin{figure}[htb]
  \includegraphics[width=8cm]{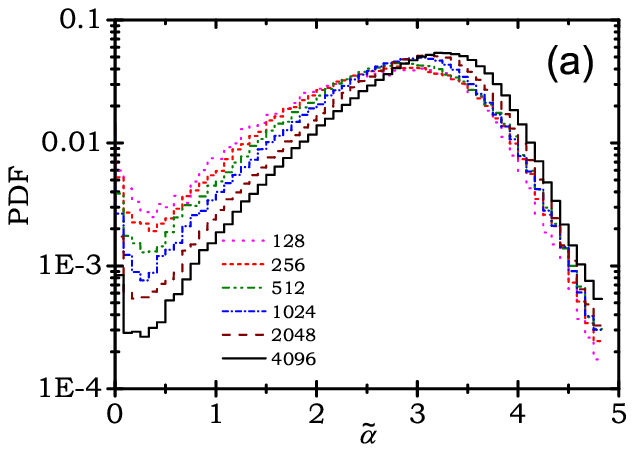}\\
  \includegraphics[width=8cm]{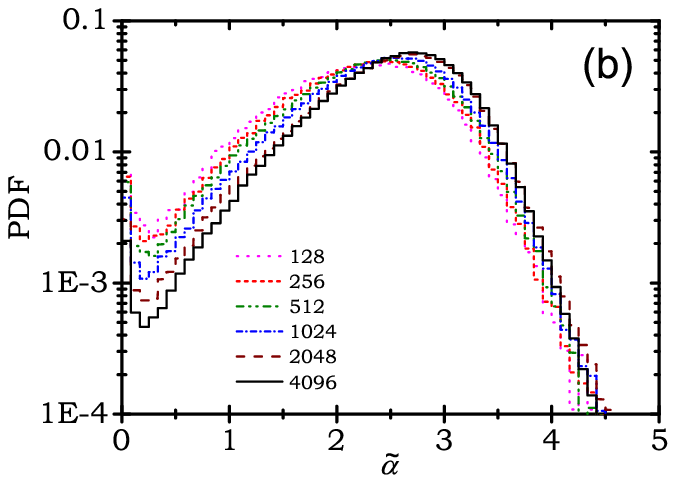}\\
  \includegraphics[width=8cm]{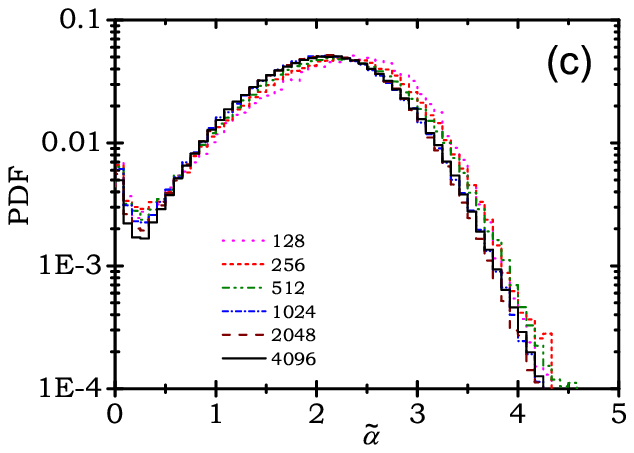}
  \caption{\label{fig:pdfte}Same as Fig.~\ref{fig:pdftm} for
  TEM eigenmodes.
  For each figure 12\% of the eigenmodes are used with (a) most positive,
  (b) closest to zero, and (c) most negative real part of their eigenvalues.
  Critical scaling is only observed in (c).
  The shift in the peak of the distribution shows that the
  rest of the eigenmodes are localized.}
\end{figure}

In the regime of weak
coupling, $Wk>10$, the numerical results show convincing indication of critical
scaling in a band of the TEM modes. These results are plotted in Fig.~\ref{fig:pdfte}.
The band of critical modes consists of those with the
most negative real part of their eigenvalues. Outside
this band the eigenfunctions
show scaling behavior similar to localized modes as shown in
Fig.~\ref{fig:pdfte}(a) and (b).
This crossover from localized to critical eigenfunctions
may be useful for measuring the critical exponent. However,
critical exponent must be defined based on a proper ordering
of eigenvalues, which is known to be a non-trivial task
for complex eigenvalues.

\subsubsection{TEM modes in the strong coupling
regime}\label{sec:strongcoupling}
\begin{figure}[htb]
  \includegraphics[width=8cm]{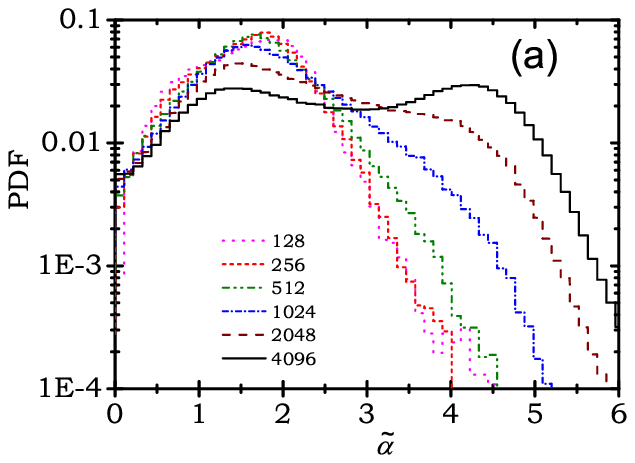}\\
  \includegraphics[width=8cm]{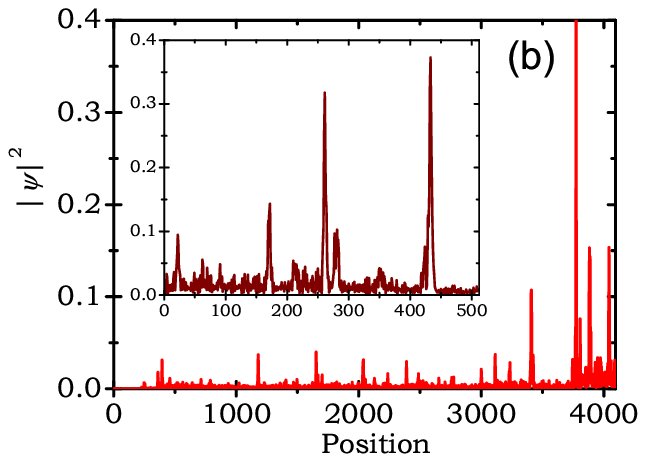}\\
  \caption{\label{fig:pdfstrong}(a) Similar to Fig.~\ref{fig:pdftm}
  for TEM modes in the regime of strong coupling with $k=1$ and
  $W=0.7$. The analyzed eigenmodes are selected from the spectral
  region indicated
  by the dashed triangle in Fig.~\ref{fig:spectrate}(b). (b) Typical
  eigenmodes used for the PDF in (a) for two system lengths
  $L=4096$ and $L=512$ (inset).}
\end{figure}

In the strong coupling regime, i.e. weak disorder, we have found it more
representative to order the complex eigenvalues by their argument.
A narrow region near the negative real axis is selected as
shown in Fig.~\ref{fig:spectrate}(b).
The histograms representing the scaled PDF of these
eigenfunctions are plotted in Fig.~\ref{fig:pdfstrong}(a).
These histograms do not overlap so the criticality cannot be
verified. Meanwhile, the behavior is
neither representative of the localized modes nor the extended modes.
It appears that the overall extent of the state is comparable
with the system size even for the longest chain, but it has
an strongly fluctuating internal structure, similar to
critical states. Typical eigenmodes of this regime are
shown
in Fig.~\ref{fig:pdfstrong}(b). Since the scaled PDF
histograms
do not overlap, we cannot prove the multifractal nature
of the states with a formal logarithmic scaling.
Describing the true nature of these modes and
their statistical behavior needs further theoretical
modeling.

\subsubsection{PDF of the intermediate models}
The results of the perturbation calculations in Sec.~\ref{sec:perturbation} are
insensitive to the details of the model. Therefore they cannot describe some
of our observations that are based on direct numerical
diagonalization. For example, according to the perturbation
theory, all the TEM modes in the weak coupling regime must
be critical. This prediction does not agree with the
simulation results since PDF scaling in observed for
only part of these modes. The same numerical analysis on
Hermitian random banded matrices perfectly matches the
results of perturbation results.

\begin{figure}[htb]
  \includegraphics[width=8cm]{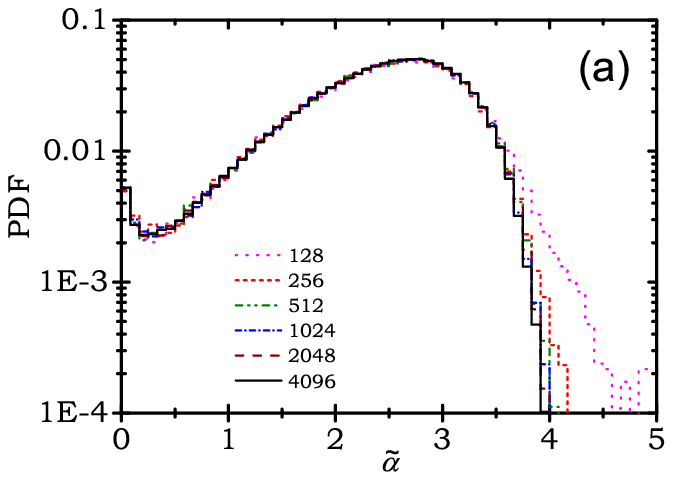}
  \includegraphics[width=8cm]{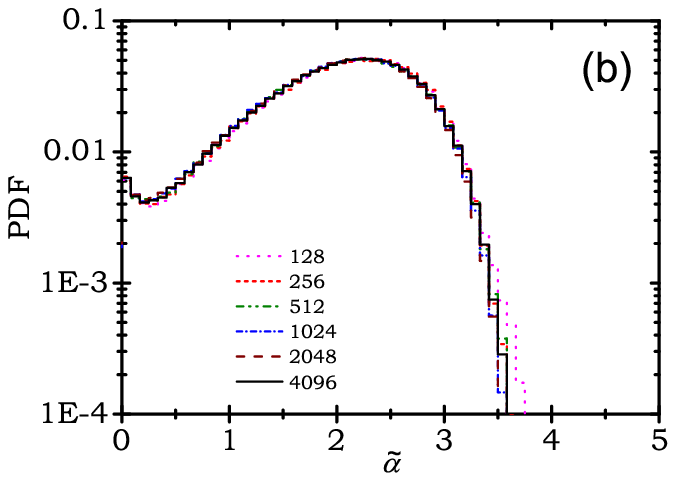}\\
  \includegraphics[width=8cm]{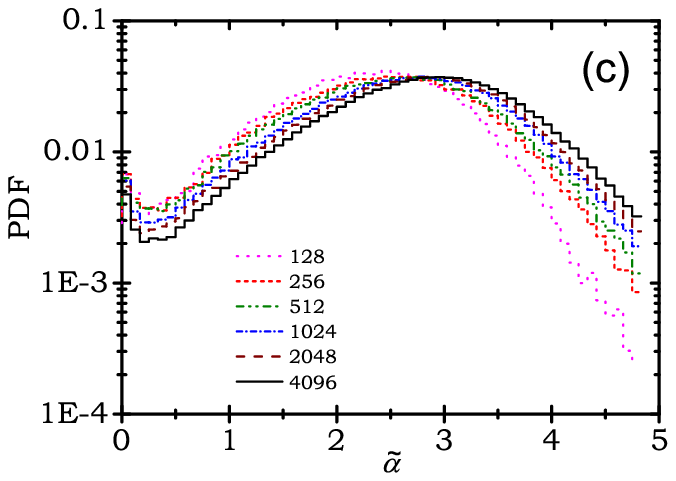}
  \includegraphics[width=8cm]{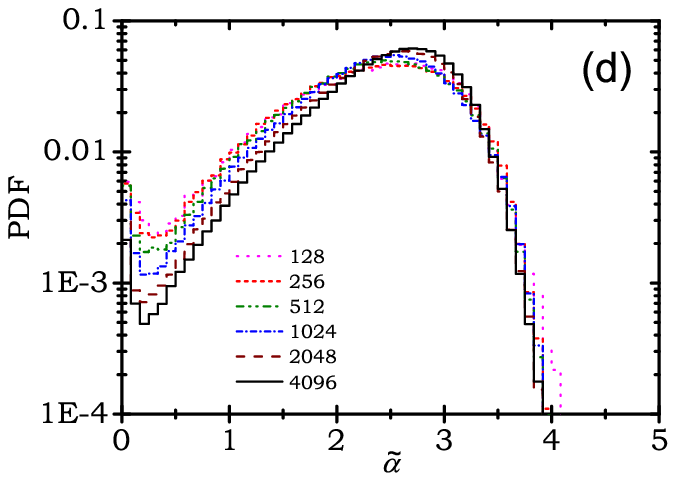}
  \caption{\label{fig:pdfhyp}Similar to Fig.~\ref{fig:pdftm}
  for models (a) H0, (b) H1, (c) H2, and (d) C1. Critical scaling
  is observed for H0 and H1 models. The eigenvectors of
  the H2 and C1 models are localized. These models are defined in
  Sec.~\ref{sec:models}.}
\end{figure}

To further explore the origin of
this deviation for complex-symmetric matrices of our model
for TEM excitations, we have
performed the same numerical procedures on the hypothetic models
introduced in Sec.~\ref{sec:models}. The PDF scaling graphs for these
models are depicted in Fig.~\ref{fig:pdfhyp}. All these
results are for the regime of weak coupling with the same
$W$ and $k$.
\begin{enumerate}
\item[H0,] The matrices in this model are orthogonal and
they are the closest to the frequently used PLRBM ensemble
with an
interaction decay exponent $\mu=1$. The PDF shows
perfect scaling as depicted in Fig.~\ref{fig:pdfhyp}(a).
The statistics is obtained by sampling from 12\% of the
eigenvectors at the band center, with eigenvalues closest
to zero. The analysis shows the same critical behavior
(not shown) for the two
ends of the spectrum. These results also confirm that our
choice of numerical precision and sampling is sufficient
for the essential conclusions we get.
\item[H1,] These matrices are also Hermitian like model H0.
The magnitude of the off-diagonal elements is not random, but
follows the decay profile of TEM complex-valued coupling~(\ref{eq:dimlesste}).
Only the phase is
randomized. Critical scaling of the eigenfunctions is again evident from the PDF
scaling depicted in Fig.~\ref{fig:pdfhyp}(b).
\item[C1,] This ensemble of complex-symmetric matrices resembles the TEM model.
The phases of the off-diagonal elements are randomized like the model H1.
The finite-size scaling of PDF, depicted
in Fig.~\ref{fig:pdfhyp}(c) shows the behavior that is
attributed to localized modes. For localized eigenvectors
the peak of the distribution shifts toward higher values
of $\alpha$, which signifies a higher density for points
with a low intensity.
\item[H2,] This model is the Hermitian form of
TEM coupling matrix. The difference between this model and H1 is
in the phase factor, which is kept periodic like the original Green function.
The only random elements of these matrices are the diagonal ones.
Despite the minor difference between models H1 and H2, the
result of PDF scaling analysis is completely different.
These results are depicted in Fig.~\ref{fig:pdfhyp}(d) and
show that the eigenvectors are localized. This observation
is inconsistent with the perturbation theory, which
predicts critical behavior for this model like H0 and H1.
It is important to point out that the considered periodicity
for the interaction phase $k=1$ is incommensurate with the
periodicity of the lattice, which equals $2\pi$ in our redefinition of units.
\end{enumerate}

\subsection{Multifractal analysis}
Since the critical scaling of part of the TEM eigenmodes in
the weak coupling regime is clearly observed in the scaling of
PDF, we apply generic techniques of multifractal(MF) analysis
to quantify the MF spectrum and compare it with our
theoretical results.

\begin{figure}[htb]
  \includegraphics[width=8cm]{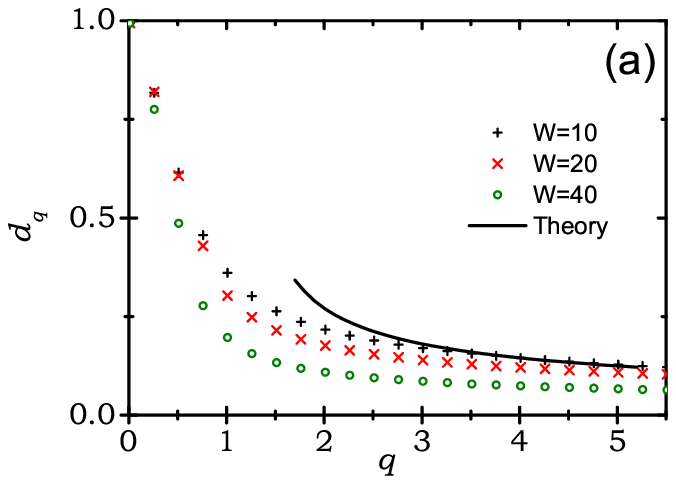}
  \includegraphics[width=8cm]{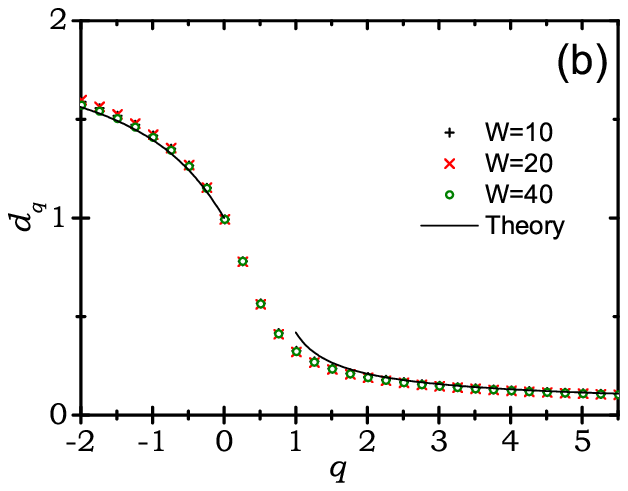}
  \caption{\label{fig:dqvq}The anomalous dimensions $d_q$ (symbols)
  at a fixed coupling strength for (a) TEM critical modes and (b) Hermitian matrices H1
  are compared with the corresponding results (solid line)
  (\ref{eq:dq})~and~(\ref{eq:dqhermitian}) from perturbation analysis. For
  Hermitian matrices, the symmetry relation predicted in
  Ref.~[\onlinecite{mirlin_exact_2006}]
  is used for plotting the theoretical curve at negative $q$. The errors estimated from the least squares fitting routine are smaller than the symbol sizes and are not shown. The largest error in $d_q$ for point $q=5.5$ on the graph is $\pm0.02$.}
\end{figure}

\begin{figure}[htb]
  \includegraphics[width=8cm]{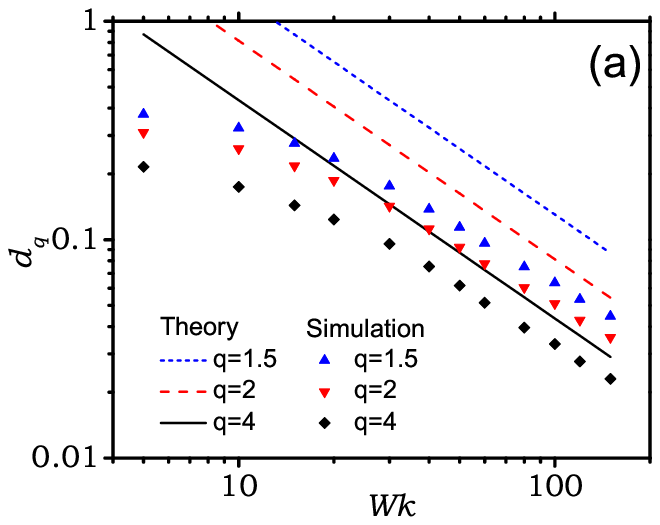}
  \includegraphics[width=8cm]{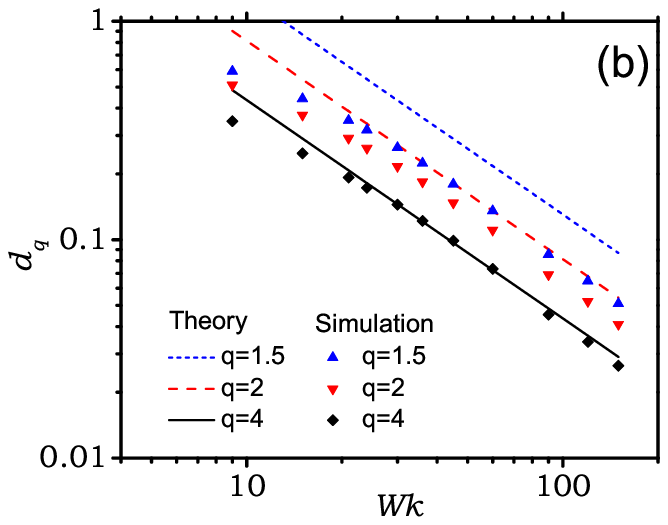}
  \caption{\label{fig:dqwk}The anomalous dimensions $d_q$
  extracted from direct numerical diagonalization (symbols)
  for three different values of
  $q$ are compared with the perturbation results (lines) of
  Eq.~(\ref{eq:dq}) in the weak coupling regime
  for different values of disorder and (a) $k=1$ or (b) $k=3$.
  The numerical results converge to the theory slowly.}
\end{figure}

We have used both size scaling and box scaling methods for
extracting the scaling exponents of GIPR for several different
parameters.
We do not observe significant differences in the results of
either method (comparison not shown).
Therefore, due to its faster computation, we use the box
scaling analysis on the largest system sizes, $L=4096$, to
extract the anomalous exponents $d_q$ for several values of
$W$~and~$k$. A summary of these results for different values of
disorder strength is depicted in
Figures~\ref{fig:dqvq} and \ref{fig:dqwk}.
To show the precision of the numerical
analysis, we have also performed this analysis for Hermitian
model H1. The results are shown in Fig.~\ref{fig:dqvq}(b)
and compared with the theoretical
prediction of Eq.~(\ref{eq:dqhermitian}). Excellent matching
between theory and simulation is evident for the Hermitian
case. However, for the complex-symmetric matrices
(corresponding to TEM coupling) the numerical results show
significant deviations from the prediction of perturbation
analysis, indicating that the first order
virial expansion is insufficient for describing that model.

In particular, according to Eq.(\ref{eq:dq}) $d_q$ is proportional
to the coupling strength $g\equiv(Wk)^{-1}$  in the weak coupling regime.
The results of direct diagonalization show, in contrast, a dependence of
$d_q$ on $W$ at a fixed value of $g$. This fact can be seen
 in Fig.~\ref{fig:dqvq}(a). The numerical results are systematically lower
than the theoretical prediction for $k<3$.

Furthermore, the dependence of MF dimensions on the coupling
strength is checked for $9\leq Wk \leq 150$. The results are shown in
Fig.~\ref{fig:dqwk} for $k=1$ and $k=3$. The overall inverse linear behavior
is observed for $Wk>30$. But the quantitative correspondence
between the numerical results and prediction~(\ref{eq:dq})
from perturbation analysis is only met for the large values of $k$
and high moments of GIPR, $q>3$.

\subsection{The singularity spectrum}
Another frequently used representation of the multifractality
is called the singularity spectrum, $\fa$. This representation is
completely analogous to the one using anomalous dimensions. For the sake of completeness,
we also show this representation for two of the models that are critical in their scaling behavior.

The singularity spectrum $\fa$ is the
fractal dimension of the subset of those
points in the wavefunction for which
intensity scales as
$L^{-\alpha}$.
It is related to the set of anomalous
exponents $\tauq \equiv d_q(q-1)$ by a Legendre
transform
\begin{equation}\label{eq:ltransform}
\fa=q\alpha-\tauq,\;\;\;\;\alpha=\taupq\;\;\;q=\fpa,.
\end{equation}
The quantity $\alpha$ introduced here is related by an irrelevant scaling prefactor to $\tilde{\alpha}$,
which was used for the definition of PDF.
However, specially for a skewed PDF, this prefactor can significantly deviate from
unity and therefore
$\alpha$ and $\tilde{\alpha}$ are not exactly equivalent quantities.

For a precise derivation of $\alpha$ and $\fa$ one has to perform a full scaling analysis on the intensity distribution. This is either possible by applying relation~(\ref{eq:ltransform}) to the calculated set of anomalous exponents or by a direct processing of the wavefunction intensities. The latter method, which was introduced by Chhabra and Jensen~\cite{chhabra_direct_1989}, is computationally superior. For this method there is no need for a Legendre transform, which is very sensitive to the numerical uncertainties.

\begin{figure}[htb]
  \includegraphics[width=8cm]{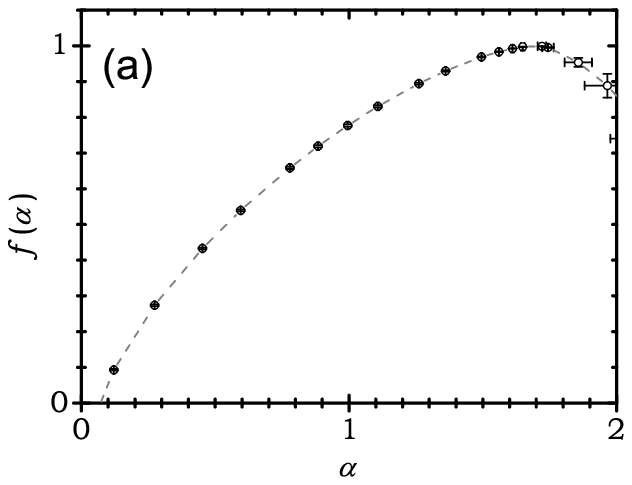}
  \includegraphics[width=8cm]{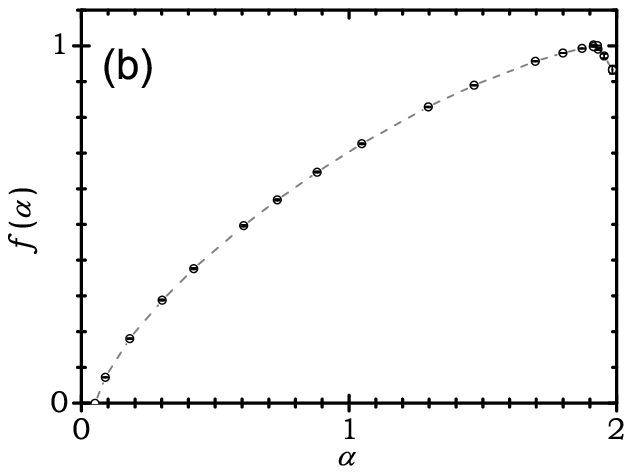}
  \caption{\label{fig:mfspec}The multifractal spectrum $f(\alpha)$
  for (a) TEM critical modes and (b) Hermitian matrices H0
  extracted directly from the eigenvectors by using the method of
  Chhabra and Jensen~\cite{chhabra_direct_1989}. For both graphs $W=30$ and $k=1$. The error bars indicate to the standard deviation among 20 realizations of disorder and are smaller than the symbol size for most of the data points. The dashed lines are guides to the eye.}
\end{figure}

We have applied the direct determination method to extract the singularity spectrum for TEM critical eigenfunctions and the Hermitian model H0. The results are shown in Fig.~\ref{fig:mfspec}. As can be seen in both graphs, the position of the peak of the spectrum is different from the peak of the corresponding PDF plots, $\tilde{\alpha}_m$, which are presented in Figures~\ref{fig:pdfte}(c) and \ref{fig:pdfhyp}(a). This difference is due to the large skewness of the PDF resulted from the very weak coupling regimes that are considered in this article.

For the Hermitian critical models the domain of $\alpha$ is restricted to $(0,2d)$ due to the symmetry relation\cite{mirlin_exact_2006} of $\fa$. Yet, there is no proof that this symmetry also holds for non-Hermitian matrices. From our data, it seems plausible that this symmetry is actually broken and there are some points with $\alpha>2$. However, to provide a strong numerical evidence for this statement one has to analyze much larger ensembles with higher numerical precision, which is beyond the scope of the current article.

\section{Summary and Conclusion}
We have investigated, theoretically and numerically, the
statistical properties of the eigenmodes of a class of
complex-symmetric random matrices, which describe the
electromagnetic propagation of polarization waves in a chain of resonant
scatterers. We have found that all of the TM modes are
localized in the weak coupling regime. The TEM modes in this regime
show critical behavior
due to the $r^{-1}$ dependence in the dyadic Green function.
This critical behavior is in agreement with the results of
the method of virial expansion for almost diagonal
matrices. We have used this method to calculate the MF
spectrum of TEM modes.

\begin{figure}[htb]
  \includegraphics[width=8cm]{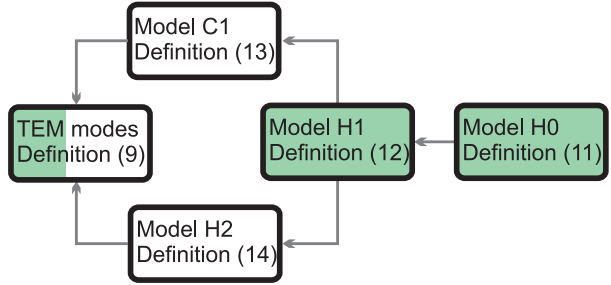}
  \caption{\label{fig:chart}Summary of the scaling analysis
  on PDF of eigenvectors of matrices from various models.
  The intermediate hypothetic models transform the Hermitian RBM to
  the model describing TEM coupling. The colored boxes
  indicate those models that show critical scaling
  behavior. The model indicated by white boxes have
  localized eigenvectors. For TEM modes, only a part of the
  spectrum is critical.
  }
\end{figure}

Although the perturbation theory suggests criticality for all TEM modes,
the numerical analysis shows this type of scaling only for
part of the spectrum in the complex plain. This is
understandable in the sense that the first oder result
of the perturbative approach gives an oversimplified picture,
which is insensitive to details of the model such as
a non-trivial phase dependence of the matrix elements.
To reveal which aspect of the TEM coupling accounts for the existence
of a critical band in the spectrum, we have analyzed three intermediate models.
These models have properties between the dipole chain
interaction matrix and power-law Hermitian banded random
matrices. The summary of the scaling results for all these
models is shown in Fig.~\ref{fig:chart}. It seems that both
non-Hermitian character of the TEM coupling and the
periodic phase of the interaction between dipoles is
important for the observed critical eigenmodes.

Our analysis also resulted in another unexpected finding.
The eigenvectors of Hermitian banded matrices with $r^{-1}$ coupling are no
longer critically scaling if the interaction phase is set
periodically. In our model H2, the randomness is only on
the diagonal. Based on the PDF scaling
results, we clearly see that the eigenvectors are localized. This is in
contrast with the commonly believed conjecture that an
interaction potential with a phase that is incommensurate
with the lattice can be considered as random.

Criticality of wavefunctions has been studied theoretically
and numerically for several models in the context of condensed matter physics.
Recently, such wavefunctions have been observed near the Anderson transition
for elastic waves\cite{faez_observation_2009} and electronic density of states at an
interface\cite{richardella_visualizing_2010}.
The recent advances in optical and microwave
instrumentation makes it possible to experiment in details
the propagation of electromagnetic waves in artificially made structures.
Our report points out to those systems in which such
critical phenomena can be directly measured.
These measurements provide a lot of insight for generic models
of wave transport in disordered system.

\section{Acknowledgements}
We thank Yan Fyodorov, Femius Koenderink, and Eugene Bogomolny
for fruitful discussions.
This work is part of the research program of the ``Stichting voor Fundamenteel
Onderzoek der Materie'', which is
financially supported by the
``Nederlandse Organisatie voor
Wetenschappelijk Onderzoek''.
AO acknowledges support from the Engineering and Physical Sciences Research Council
[grant number EP/G055769/1].

\newpage

\begin{appendix}

\section{Perturbation results for eigenvector moments}\label{app:dq}
In the regime of the strong multifractality, i.e. when
$Wk\gg 1$, the density of states is determined only by the
distribution of the diagonal elements and hence is uniform.
Therefore we assume that all eigenvectors are characterized
by the same scaling exponents and we define the ensemble average
GIPR as
\begin{eqnarray}\label{mom_def}
\langle P_q\rangle&=&\frac{1}{L}\sum_{n=1}^L \sum_{i=1}^L
|\psi_n(x_i)|^{2q}.
\end{eqnarray}
The moments
of the eigenvectors can be extracted from the powers of the
diagonal elements of the Green functions and therefore
can be computed using the method of the the virial
expansion.

In the first order of the virial expansion, which
corresponds to a pure diagonal matrix, the eigenvectors
consists of only one non-zero component and therefore all
moments are equal to one due to normalization:
$\langle P_q\rangle^{(1)}=1$.

In the next order of the virial expansion the contribution
to the eigenvectors moments from all possible pairs of two
levels of the unperturbed system should be taken into
account. Thus we need to calculate the moments of the
eigenvectors of the following $2\times 2$ matrices:
\begin{eqnarray}
M(i,j)&=&\left(\begin{matrix} E_i & \myham_{ij}\\ \myham_{ji} &
E_j\end{matrix}\right),
\end{eqnarray}
Denoting by $P_q(i,j)$ the moments of the second component
of the corresponding eigenvectors we have
\begin{eqnarray}\label{mom_two} \langle P_q\rangle^{(2)}=\frac{1}{L}\sum_{i\neq
j}^L
(P_q(i,j)-1)
\end{eqnarray}

The subtraction of $1$ in the equation
above eliminates the contribution already
taken into account in the diagonal approximation.

Let us introduce the following notation for $M(i.j)$:
\begin{eqnarray}\label{twom}
H(i,j)&=&\left(\begin{matrix} E_1 & h e^{\imath\phi}\\
he^{\imath\phi} & E_2
\end{matrix}\right),
\end{eqnarray}
where $E_1\equiv E_i$, $E_2\equiv E_j$ and $
he^{\imath\phi}\equiv \myham_{ij}=\myham_{ji}$. Below we also denote
$P_q(i,j)$ by $P_q$. As the eigenvectors of $H(i,j)$ do not
depend on the absolute values of the matrix elements, but
only on their relative ratios, we may assume now that $E_1$
and $E_2$ are distributed uniformly in $[-1/2,1/2]$ and
$h=e^{i k |i-j|}/W k |i-j|$. For the reasons which
will be described later, the other terms in the coupling
elements are not considered here. By the assumption of weak
coupling $|h|\ll 1$ and writing the eigenvectors of $M(i,j)$ explicitly, we obtain:
\begin{eqnarray}\label{P_q}
P_q&=&\int_{-1/2}^{1/2}dE_1\int_{-1/2}^{1/2}dE_2 (Q((E_1-E_2)/h)+Q((E_2-E_1)/h)),\nonumber\\
Q(x)&=& \left(\frac{4}{4+ \left(x-\sqrt{(x^2+4 e^{-2 i
\phi
   }}\right) \left(x-\sqrt{(x^2+4  e^{2 \imath \phi
   }}\right)}\right)^q.
\end{eqnarray}
In the limit $h\to 0$ one can show that $Q(r)=\theta (r)$
and thus $P_q=1$. This result corresponds to the diagonal
approximation and is cancelled by $-1$ in
Eq.(\ref{mom_two}). In order to find first non-trivial
contribution we need to compute a term which is linear in
$h$. To this end we first differentiate Eq.(\ref{P_q})
with respect to $h$, then change the integration variables
$\{E_1,E_2\}\to\{E_1,x=(E_1-E_2)/h\}$ and consider the
limit $h\to 0$:
\begin{eqnarray}
\lim_{h\to 0} \frac{d P_q}{d
h}&=&-2\int_{-\infty}^{\infty}dx\: x Q^{\prime}(x)\equiv
-F(q,\phi).
\end{eqnarray}
The expansion of $P_q$ then takes the form:
\be
P_q=1-|h|F(q,\phi)+O(h^2). \en
Collecting the
contributions from all the off-diagonal elements we obtain
\be \langle P_q\rangle=1-\frac{1}{W k}\frac{1}{L}\sum_{i\neq
j}\frac{F(q,k|i-j|)}{|i-j|}. \en
Expanding $F(q,\phi)$ in
the Fourier series $F(q,\phi)=\sum_{p}
c_p(q)e^{\imath p\phi}$ we find
\begin{eqnarray}
\frac{1}{L}\sum_{i\neq
j}\frac{F(q,k|i-j|)}{|i-j|}&=&\frac{1}{L}\sum_{i\neq
j}\frac{c_0(q)}{|i-j|}+\frac{1}{L}\sum_{i\neq j}\sum_{p
\neq 0}\frac{c_p(q)e^{\imath p k|i-j|}}{|i-j|} \nonumber\\
&=&2c_0(q)\ln L
+O(1),
\end{eqnarray}
so that we arrive at the following result:
\be
\langle P_q\rangle=1-\frac{2 c_0(q)}{W k}\ln L,\quad
c_0(q)=\frac{1}{\pi}\int_0^{2\pi}d\phi
\int_{-\infty}^{\infty}dx\: x Q_x^{\prime}(x,q,\phi), \en
where $Q$ is defined in Eq.(\ref{P_q}) and can be written
as \be Q(x,q,\phi)=\left(\frac{4}{4+ \left|x-\sqrt{x^2+4
e^{-2 i \phi}}\right|^2}\right)^q.
\en
Comparing this result with the scaling of the moments
by changing the system size
\be \langle P_q\rangle \propto L^{-d_q(q-1)},
\en we find the expressions for the fractal dimensions
\be
d_q=\frac{2 c_0(q)}{W k(q-1)}. \en
One can check that in
the case of the orthogonal matrices, i.e. when $\phi=0$, the
old result can be reproduced after the change of the
variable $x=(2w-1)/\sqrt{w(1-w)}$:
\begin{eqnarray}
c_0(q)&=&2 \int_{-\infty}^{\infty}dx\: x Q_x^{\prime}(x,q,0)\nonumber\\
&=&2\int_0^1dw\:\frac{qw^{q-3/2}(2w-1)}{\sqrt{1-w}}\nonumber\\
&=&2\sqrt{\pi} \frac{\Gamma (q-1/2)}{\Gamma (q-1)}\\
d_q(\phi=0)&=&\frac{4\sqrt{\pi}}{Wk}\frac{\Gamma
(q-1/2)}{\Gamma (q)}
\end{eqnarray}
The result for complex-symmetric matrices deviate only slightly from the
one for orthogonal case. This deviation is most pronounced
around $q=2.5$.

%
%


\end{appendix}

\end{document}